\documentclass[12pt]{JHEP3}

\usepackage{graphicx}
\usepackage{cite}
\usepackage[english]{babel}
\usepackage{amssymb}
\usepackage{amsfonts}
\usepackage{amsmath}

\newcommand*\chem[1]{\ensuremath{\mathrm{#1}}}

\relax
\def\be{\begin{equation}}
\def\ee{\end{equation}}
\def\bea{\begin{eqnarray}}
\def\eea{\end{eqnarray}}

\newcommand{\bsea}{\begin{subeqnarray}}
\newcommand{\esea}{\end{subeqnarray}}
\newbox\pippobox

\def\6{\partial}

\def\a{\alpha}

\def\nn{\nonumber}

\def\sq
\def\a{\alpha}

\def\dd{\mathrm{ d}}

\def\re{\operatorname{Re}}
\title{AC conductivity for a holographic Weyl Semimetal}

\author{Gianluca Grignani\thanks{gianluca.grignani@pg.infn.it},
	 Andrea Marini\thanks{andrea.marini@pg.infn.it}, 
	 Francisco Pe\~na-Benitez\thanks{benitez@pg.infn.it},
	 Stefano Speziali\thanks{stefano\_speziali@libero.it}
 \\

Dipartimento di Fisica e Geologia, Universit\`a di Perugia,\\
I.N.F.N. Sezione di Perugia,\\
Via Pascoli, I-06123 Perugia, Italy\\
}

\preprint{}

\abstract{
	We study the AC electrical conductivity at zero temperature in a holographic model for a Weyl semimetal. At small frequencies we observe a linear dependence in the frequency. The model shows a quantum phase transition between a topological semimetal (Weyl semimetal phase) with a non vanishing anomalous Hall conductivity and a trivial semimetal. The AC conductivity has an intermediate scaling due to the presence of a quantum critical region in the phase diagram of the system. The phase diagram is reconstructed using the scaling properties of the conductivity. We compare with the experimental data of \cite{Xu:2015vt} obtaining qualitative agreement. 
}

\keywords{Holography, Anomaly induced transport, Weyl semimetals, AC conductivity, strongly correlated system, Quantum phase transitions.}

\begin{document}

\section{Introduction}\label{intro}

The recent discovery of Weyl semimetals (WSM) \cite{Xu:2015jx} has increased the interest of both high energy and condensed matter communities in their description. WSM  are three dimensional gapless semiconductors. Their low energy excitations are massless fermions, called Weyl fermions. Due to the Nielsen-Ninomiya theorem \cite{Nielsen:1983rb} these fermions come in pairs with opposite chirality (right- and left-handed). In some sense these materials can be viewed as a 3d version of graphene. WSM are characterized by the presence of singular points in the Brillouin zone where the conduction and valence band touch. If time reversal or parity is broken, left- and right-handed quasiparticles can  sit at different points in the Brillouin zone. The fact that quasi-particles around the Weyl points obey a massless relativistic equation implies astonishing properties for these materials, as chiral magnetic \cite{Kharzeev:2006wk,Kharzeev:2007tn,Gynther:2010ed} and vortical effects \cite{Erdmenger:2009ky,Banerjee:2011kt,Landsteiner:2011cp,Landsteiner:2011iq}, odd (Hall) viscosity \cite{Landsteiner:2016stv}, negative magnetoresistivity \cite{Landsteiner:2014vua,Lucas:2016omy,Li:2014bha}. All these phenomena are intimately related with the quantum  anomalies observed in massless fermionic systems. Another characteristic property of WSM is the presence of edge states, called Fermi arcs \cite{Batabyal:2016ek,Xu:2015jx}.

As it happens in graphene, the Fermi velocity of the electrons in WSM is at least two orders of magnitude smaller than the speed of light \cite{Gonzalez:2015tz}. Therefore the analogue of the fine structure constant can be a hundred times bigger than its QED counterpart, opening the possibility of the strong coupling regime. However, most of the attempts to describe WSM have been from a weakly coupled perspective \cite{Burkov:2011de,Roy:2016amv,Hosur:2011fl,Hosur:2013eb}. From a strong coupling point of view some effort has also been done \cite{Gonzalez:2015tz,Gursoy:2012ie,Jacobs:2015fiv,Landsteiner:2015lsa,Landsteiner:2015pdh}.

At zero temperature Weyl materials are supposed to show quantum phase transitions connecting quantum anomalous Hall insulators, insulators and the Weyl semimetal phases \cite{Burkov:2011de}.

Inspired by the recent experiment\footnote{We acknowledge Maria A. H. Vozmediano for letting  us know about the existence of this paper.} \cite{Xu:2015vt}, we studied the optical response of a strongly coupled Weyl semimetal using holography. From a holographic point of view, the problem of the optical conductivity has been already addressed in the past \cite{Jacobs:2015fiv}, however  the aspects of having a splitting of Weyl cones was ignored, also the presence of the chiral anomaly was not implemented. For these reasons, we choose the perspective of \cite{Landsteiner:2015lsa}, in which separation of the Weyl cones and the axial anomaly are effectively implemented.

The observations of  \cite{Xu:2015vt} are partially consistent with the weak coupling predictions of \cite{Hosur:2011fl,Burkov:2011de}. For frequencies higher than the temperature of the material, the conductivity is characterized by a straight line that changes its slope at some energy scale. The first slope is consistent with the prediction \cite{Hosur:2011fl}
\be\label{eq:condweak}
\sigma \approx n\frac{e^2}{12h}\frac{\omega}{v_f},
\ee
in which $n=8$ is the number of Weyl points around the Fermi energy. They used this expression to fit the value of the Fermi velocity, obtaining a value consistent with previous predictions for this material. The second slope is interpreted as a changing in number of degrees of freedom due to the presumable influence of the rest of Weyl points at such energies. Nonetheless in this case the Fermi velocity obtained does not lie within a reasonable value. Certainly there could be many effects influencing the intermediate behavior of the conductivity, however the robust persistence of a power law at those energies suggests that the physics of the Weyl cones is still determining the optical response. An alternative explanation implied by our toy model  could be given by the presence of the quantum critical point and its corresponding  quantum critical region. 

The rest of the paper is organized as follows. In section \ref{sec:Themodel} we review and generalize the model of \cite{Landsteiner:2015lsa}. Then, in section \ref{sec:Phases}, we classify the IR geometries, which are dual to the ground states of the system. In section \ref{sec:Conduc} we compute analytically the IR frequency dependence of the conductivity for the different phases of the model, and numerically the full frequency dependence of the conductivity matrix. We also reconstruct the quantum phase diagram of the model using the power law properties of the conductivity. In section \ref{sec:expe} we compare our results with \cite{Xu:2015vt} and we finish with our conclusions.

{\bf Note added:} When finishing the paper we received the manuscript \cite{Copetti:2016wz} which shows some overlap with our work. They studied the anomalous Hall conductivity in the axial current and considered a top down model.

\section{Holographic effective theories for a Weyl Semimetal}
\label{sec:Themodel}

A holographic model for Weyl semimetals has been proposed in \cite{Landsteiner:2015lsa} and studied in more detail in \cite{Landsteiner:2015pdh} and \cite{Landsteiner:2016stv}. The main features of this model are the implementation of the chiral anomaly and effective separation of Weyl cones. The model undergoes a quantum phase transition from a topologically non-trivial semimetal to a trivial one. Previous holographic models \cite{Jacobs:2015fiv,Jacobs:2014nia, Gursoy:2012ie} have considered a semi-holographic point of view, coupling weakly coupled Weyl fermions to a strongly coupled quantum critical system. 

We shall follow the spirit of  \cite{Landsteiner:2015lsa} to implement a time-reversal breaking parameter\footnote{Effective separation of Weyl cones.} and a ``mass operator", as deformations of a strongly coupled conformal field theory. The holographic action we propose is a generalization of the action used in \cite{Landsteiner:2015lsa}
\bea\label{eq:action}
S &=& S_0 + S_{GH} + S_{CT}\\ 
\nn S_0 &=& \int \dd^5x \sqrt{-g} \left[ \, R- \mathcal V(|\phi|)  - \frac{1}{4} Z_1(|\phi|)H^2 - \frac{1}{4}Z_2(|\phi|)F^2 - Z_3(|\phi|)|D \phi|^2 \right.\\
&&+ \left.\frac{\alpha}{3}\epsilon^{M N R P Q} A_{M} \big( F_{NR} F_{PQ} + 3 H_{NR}H_{PQ}\big)\right] \, ,
\eea
where $\epsilon_{MNRPQ}=\sqrt{-g}\varepsilon_{MNRPQ}$ and $\varepsilon_{0123r}=1$.  $S_{GH}$ and $S_{CT}$  are the Gibbons-Hakwing boundary action and the counterterm that renormalizes the on-shell action, respectively. We show them in appendix \ref{app:CT}.
 
In Holography, currents associated to global symmetries of the QFT are dual to gauge fields propagating in the five dimensional geometry, whereas quantum anomalies are implemented by Chern Simons terms in the action \cite{Witten:346620}.  Let us explain the ingredients of the model:
\begin{itemize}
\item The field dual to the vector (electromagnetic) boundary current is $V_M$ and $H_{MN}=\partial_M V_N - \partial_N V_M$  its corresponding field strength. The asymptotic value of $V_M$ is the source for the electromagnetic current, i.e. it corresponds to a non dynamical background electromagnetic field switched on in the dual field theory.
\item $A_M$ is the field dual to the (anomalous) axial current and its corresponding field strength is $F_{MN}=\partial_M A_N - \partial_N A_M$. The boundary value of $A_M$ is the source for the anomalous axial current. 
\item The Chern Simons coupling has been properly tuned in order to reproduce the Ward's identities of the {\it consistent} (conserved) vector and (anomalous) axial currents of the boundary QFT \cite{Landsteiner:2015lsa}
\bea
\partial_\mu J^\mu &=& 0\,,\\
\label{eq:wardj5}\partial_\mu J_5^\mu &=& -\frac{\alpha}{3}\varepsilon^{\mu\nu\rho\lambda}\left(F_{\mu\nu}F_{\rho\lambda} + 3 H_{\mu\nu}H_{\rho\lambda}\right) \ldots
\eea
where dots refer to the contribution coming from the explicit breaking of $U(1)_{axial}$ given by the presence of the mass deformation. Comparing the Ward identity for $N$ Dirac fermions with Eq. (\ref{eq:wardj5}) we can fix the Chern-Simons coupling to be
\be
\alpha = \frac{N}{16\pi^2},
\ee
$N$ is the flavor number.
\item In order to have a conformal UV fixed point, i.e. an asymptotically AdS space the scalar potential has to satisfy
\be
\mathcal V_{UV} \equiv \mathcal V(|\phi_{UV}|)=-\frac{12}{L^2}\,.
\ee
\item The holographic dictionary establishes that the scalar field $\phi$ is dual to a certain scalar operator $\mathcal{O}$ with scaling dimension $[\mathcal{O}]=\Delta_\phi$ given by
\be
\Delta_\phi = 2+ \sqrt{4 + (m L)^2}\,,
\ee
where $m^2 = 1/2\mathcal V''_{UV}$. If we choose $(mL)^2=-3$,\footnote{Notice that AdS BF bound $(mL)^2>-4$ is not violated for this mass.} the operator $\mathcal{O}$ will have scaling dimension $\Delta_\phi=3$ and necessarily will couple to a source $M$ of dimension $[M]=1$. So we are allowed to interpret the scalar field as a mass deformation of the CFT.

\item  The scalar field is charged only under the axial field, and the covariant derivative reads $D_M=\partial_M -iqA_M$. Notice that if the scalar field were charged under the vector field $V_M$ the electromagnetic current would not be conserved.
\end{itemize}

Writing the scalar field as $\phi=\psi e^{i\theta}$, the gauge invariance of the system allows us to set $\theta=0$. In this way the action can be written as follows
\bea\label{eq:action2}
\nn S_0 &=& \int \dd^5x \sqrt{-g} \left[ \, R- \mathcal V(\chi)  - \frac{1}{4} Z_1(\chi)H^2 - \frac{1}{4}Z_2(\chi)F^2 - (\partial \chi)^2 + W(\chi)A^2\right.\\
&& \left.+\frac{\alpha}{3}\epsilon^{M N R P Q} A_{M} \big( F_{NR} F_{PQ} + 3 H_{NR}H_{PQ}\big)\right] \, ,
\eea
where the new scalar field is defined by $\partial_\chi \psi(\chi) = (Z_3(\chi))^{-1/2}$ and $W(\chi) = q^2\psi^2Z_3$. 

Considering that we already proved that (\ref{eq:action}) is contained  in (\ref{eq:action2}), we will use (\ref{eq:action2}) as the fundamental action without assuming any specific form for the functions $\mathcal V,Z_1,Z_2,W$.
The holographic dictionary establishes that the on-shell gravity action corresponds to the generating functional of the QFT. Therefore taking variations of the action with respect to the sources we obtain the one-point functions of the corresponding associated operators

\be
J^\mu = \frac{\delta S}{\delta v_\mu}  \qquad , \qquad J_5^\mu = \frac{\delta S}{\delta b_\mu} \qquad , \qquad \mathcal{O} = \frac{\delta S}{\delta M}\,,
\ee
where $v_\mu= V_\mu(\Lambda)$,  $b_\mu= \Lambda^{-\Delta_b}A_\mu(\Lambda)$,\footnote{The scaling dimension of $J_5$ is $[J_5]=3+\Delta_b$, with $\Delta_b = -1 + \sqrt{1 + \frac{2W_{UV}L^2}{ Z_2(UV)}}$. Notice that in the case of the abelian Higgs action (\ref{eq:action}) $\Delta_b=0$. For simplicity, and assuring that the explicit $U(1)_{axial}$ breaking is only due to the presence of the scalar mass operator we will assume $W(\chi_{UV})=0$. We will also assume $Z_1(\chi_{UV})=Z_2(\chi_{UV})=1$.} and $M= \Lambda^{4-\Delta_\phi}\chi(\Lambda)$ are the field theory sources. $\Lambda$ is the UV radial cutoff where we put the field theory to live before renormalising and sending it to infinity.

Taking variations of the renormalized action (\ref{eq:action2}), we obtain the consistent currents and scalar operator 
\bea
\label{eq:conscurrV}
J^{\mu} &=& \lim_{\Lambda\to\infty}\bigg[ \sqrt{-g}Z_1 H^{\mu r} + 4 \alpha \varepsilon^{\mu \nu \rho \lambda} A_{\nu} H_{\rho \lambda} \bigg]_{\Lambda}+ \frac{\delta S_{CT}}{\delta v_\mu}\,,\\
J^{\mu}_5 &=& \lim_{\Lambda\to\infty}\bigg[ \sqrt{-g}Z_2 F^{\mu r} + \frac{4}{3} \alpha \varepsilon^{\mu \nu \rho \lambda} A_{\nu} F_{\rho \lambda} \bigg]_{\Lambda} + \frac{\delta S_{CT}}{\delta b_\mu}\, ,\\
\mathcal{O} &=&  -2\lim_{\Lambda\to\infty}\Lambda^{\Delta_\phi-4} \sqrt{-\gamma} \partial_r \chi \bigg|_{\Lambda} + \frac{\delta S_{CT}}{\delta M}\,.
\eea

\section{IR fixed points}
\label{sec:Phases}

As it is well known, this type of theories may have an RG flow from the UV AdS fixed point to other scaling fixed points in the IR, where the flow is tuned by the running of the scalar field \cite{Charmousis:2010zz,Gubser:2008wz,Bhattacharya:2014uu,Gouteraux:2012jw,Gath:2012pg}. In our analysis we will consider the  case where the scalar runs to a constant. On top of that, the authors of \cite{Landsteiner:2015pdh} proved that the IR behavior of the solution determines whether the system is in a Weyl semimetal or in a trivial semimetal phase, depending on whether the spatial component of the axial gauge field runs to a constant or to zero. Restricting to the case of neutral zero temperature IR geometries we  propose the following ansatz
\bea
\label{eq:ansatz}
\dd s^2 &=& u(r)(-\dd t^2 + \dd x_1^2+\dd x_2^2)  + h(r)\dd x_3^2 +\frac{\dd r^2}{u(r)}\,,\\
A &=& A_3(r)\dd x_3 \,,\\
 \chi &=& \chi(r) \,,
\eea
The equations of motion of the system are shown in the appendix~\ref{app:EOM}. As previously pointed out, we look for scaling IR geometries with constant scalar field at the leading order,  \mbox{$\chi(r)=\chi_{IR}+\delta\chi(r)$}. Therefore we assume the following form for the IR metric components\footnote{Using this radial coordinate the IR region is located at $r\to 0$.}
\be
\label{eq:ansatzIR} u(r) = u_0 r^{2}\left(1+\delta u(r)\right) \qquad,\qquad h(r) = h_0 r^{2\beta}\left(1+\delta h(r)\right) \,,
\ee
together with the consistency conditions\footnote{Notice that we have included subleading corrections to the IR fields; knowing the form of the corrections is necessary to understand the reliability of the IR solution.}
\be
u_0>0 \,, \qquad h_0>0\,,\qquad Z_2^{IR}>0\, .
\ee

With these assumptions, the leading order IR Einstein's equations take the form
\bea\label{eq:eins1IR}
\frac{W_{IR}}{2 h_0 u_0} \left(\frac{A_3}{r^{  \beta }}\right)^2 - \frac{  Z_2^{IR}}{4 h_0}\left(\frac{A'_3}{r^{\beta-1 }}\right)^2 +\frac{ \mathcal V_{IR}}{2 u_0}+3 (\beta +1)  &=& 0\,,\\
\label{eq:eins2IR}\frac{  Z_2^{IR}}{2 h_0}\left(\frac{A'_3}{r^{\beta-1 }}\right)^2- (1-\beta )\beta   &=& 0\,,\\
\label{eq:eins3IR}\frac{ Z_2^{IR}}{2 h_0}\left(\frac{A'_3}{r^{\beta-1 }}\right)^2 -\frac{ W_{IR}}{ h_0 u_0} \left(\frac{A_3}{r^{  \beta }}\right)^2 - (1-\beta ) (\beta +3) &=& 0 \,.
\eea
These equations can be consistently solved for two possible cases:

\begin{itemize}
	\item {\bf Critical or Marginal solution:} If $A_3(r)= r^\beta\left(1+\delta A_3(r)\right)$ the equations become an algebraic system that can be solved exactly for $u_0$ and $h_0$
\be
u_0 = -\frac{\mathcal V_{IR}}{9+ \beta  (\beta +2)} \quad , \quad  h_0 =  \frac{\beta   }{2(1- \beta) }Z_2^{IR}\,.
\ee
The Maxwell and scalar equations reduce to
\be
W_{IR} = \frac{3}{2}\beta Z_2^{IR}u_0 \quad , \quad \partial_\chi \log \left( \mathcal VW^{3p} Z^{\beta p}\right)\left|_{\chi=\chi_{IR}}\right.=0\,,
\ee
with  $p=\frac{\beta -1}{\beta ^2+2 \beta +9}$, which can be used to solve for $\beta$ and $\chi_{IR}$. The consistency conditions imply 
\be
 \mathcal V_{IR} < 0 \qquad,\qquad 0<\beta <1\,.
\ee

See appendix \ref{app:IRPert} for an analysis of perturbations around the IR geometries.

	\item {\bf Irrelevant solutions:} The second possibility corresponds to the gauge field decaying in the IR fast enough such that $r^{-  \beta }A_3$ is subleading in the IR. They can be split in two disconnected cases, depending on whether the gauge field vanishes or not in the IR.
		
\begin{itemize}
	\item {\bf Trivial semimetal:}  If the gauge field vanishes, the solution for Eqs. (\ref{eq:eins1IR}-\ref{eq:eins3IR}) is AdS, with the IR length scale given by
\be
u_0 = - \frac{1}{12}\mathcal V_{IR} = L_{IR}^{-2} \qquad , \qquad \beta=1\,.
\ee
Then, using the scalar equation, the value for $\chi_{IR}$ is fixed by minimizing $\mathcal V$ and the Maxwell equation trivializes 
\be
\mathcal V'(\chi_{IR}) = 0\qquad ,\qquad \mathcal V_{IR} < 0 .
\ee

If we consider the leading corrections to AdS that preserve $T=0$ and are irrelevant in the IR, we obtain 
\be
\delta A_3 = c_b r^{\Delta_{b_{IR}} }\qquad, \qquad\delta \chi = c_\chi r^{-4+\Delta_{\chi_{IR}}},
\ee
with the IR scaling dimension of the associated operators
\be
\label{eq:andimgaugeIR}\Delta_{b_{IR} } = -1 + \sqrt{ 1 + \frac{ 2W_{IR}  L_{IR}^2 }{ Z_2^{IR} } }\,, \qquad
\Delta_{\chi_{IR}} =2+  \sqrt{ 4 + m_{IR}^2  L_{IR}^2 }\,.
\ee
Stability of the IR background requires 
\be
W_{IR}   >0 \qquad, \qquad m_{IR}^2  L_{IR}^2 > 0 \,.
 \ee
	\item {\bf Weyl semimetal phase:} There is a last inequivalent irrelevant case. If the gauge field runs to a constant, necessarily Eqs. (\ref{eq:eins1IR}-\ref{eq:eins3IR}) imply that the gauge field must be massless in the IR
	\be
	W_{IR} = 0\,.
	\ee
	If the last is satisfied, the solution is AdS
\be
u_0 = - \frac{1}{12}\mathcal V_{IR}  \qquad , \qquad \beta=1\qquad, \qquad \mathcal V_{IR}<0\,,
\ee
the peculiarity of this solution is that the scalar field must be simultaneously an extremum  of the scalar potential and the mass function
\be
\mathcal V'(\chi_{IR}) = 0  \qquad , \qquad W'(\chi_{IR}) = 0\,,
\ee

Perturbations show an irrelevant deformation,

\be
\delta\chi = c_\chi r^{-3/2}e^{-\frac{s}{r}}\,,
\ee
with $s=\sqrt{1/2W_{IR}''h_0^{-1}} L_{IR}$ and $W_{IR}''>0$. If the last inequality was not satisfied the Weyl semimetal phase would not be reliable considering that all perturbations would be relevant.
\end{itemize}
\end{itemize}

The previous analysis allowed us to have a geometrical picture of the Weyl semimetal phase. Let us remind the reader that gauge symmeties  in the bulk correspond to global symmetries in the boundary theory, therefore the condition $	W_{IR} = 0$ means that axial gauge symmetry is restored in the IR. On the other hand, the trivial phase is characterized by a massive axial gauge field in the IR, such that gauge symmetry is explicitly broken, giving an IR anomalous dimension to the operator $J_5$ determined by (\ref{eq:andimgaugeIR}). That explains why the minimal coupling of the abelian Higgs model used by \cite{Landsteiner:2015pdh} contains the Weyl semimetal phase.


\section{Conductivities}
\label{sec:Conduc}
Before computing the full frequency dependence of the conductivity, for which it will be necessary to use numerical techniques, we will analyze the behavior at small frequencies. To do so, we introduce the following consistent set of  linear fluctuations

\begin{equation}
\delta V_i = v_i(r) \,  e^{-i \omega t}
\end{equation}
where $i$ takes values $x,y,z$. The fluctuations are decomposed into two sectors, longitudinal and transverse to the background $A_M=(0,0,0,A_3,0)$. The transverse equations read
\begin{equation}
\label{eq:eqf1}
\left(u\sqrt{h}Z_1v_c' \right)'  +   \left( \omega^2\frac{Z_1\sqrt{h}}{u}\delta_{cd}   + 8i\omega\alpha  A'_3 \epsilon_{cd} \right)v_d = 0\,,
\end{equation}
with $c,d=x,y$. These equations can be diagonalized using the helicity fields $v_\pm = v_x \pm i v_y$ 
\be\label{eq:dec1}
\left(u\sqrt{h}Z_1v_\pm' \right)'  +  \left( \omega^2\frac{Z_1\sqrt{h}}{u}  \pm 8\omega\alpha  A'_z  \right)v_\pm =0\,.
\ee
The  equation for the longitudinal field can be written as
\be
\label{eq:eqf3}  \left(\frac{u^2}{\sqrt{h}} Z_1v_z'\right)'  + \frac{\omega^2Z_1 }{\sqrt{h}}v_z    =0 \, .
\ee

These equations were studied in \cite{Landsteiner:2015pdh} to compute the DC conductivities in the case  $Z_1=Z_2=1$ and $W=q^2\chi$. In this section we shall generalize their computation and we will also obtain the leading $\omega$ behavior  of the conductivities.

\subsection{DC conductivities}
We start first by computing the DC conductivities solving Eqs. (\ref{eq:dec1}, \ref{eq:eqf3}) up to linear order in $\omega$. To do so,  we construct a perturbative solution as follows
\be
\label{eq:vpm} v(r) \approx v^{(0)}(r) + \omega\, v^{(1)} (r)+ O(\omega)^2\, .
\ee
The solution for $v^{(0)}(r)$ and $v^{(1)}(r)$ can be found in an integral form even though the explicit functions of the background fields are not know analytically. After imposing regularity of fields in the IR, we obtain the linear order solutions

\bea
\label{eq:linvz}v_z(r) &\approx & 1 + \mathcal O(\omega)^2 \\
\label{eq:linvt}v_\pm(r) &\approx & 1  \mp 8\alpha\omega\int_{r}^{\infty}\dd r' \frac{A_3(r')-A_3(0)}{u(r')\sqrt{h(r')}Z_1(\chi(r'))} + O(\omega)^2\,.
\eea
Using the definition (\ref{eq:conscurrV}) we can write the {\it consistent} current as follows
\be
\label{eq:eleccurr}
J_i = \lim_{\Lambda\to\infty}\left(-\Lambda^3\frac{i}{\omega}\frac{v'_i(\Lambda)}{v_k(\Lambda)} + 8\alpha\epsilon_{ijk}b_j -  i\omega\delta_{ik}\log \Lambda \right)(i\omega v_k(\Lambda))\,,
\ee
from which we read the conductivity. After plugging the solutions (\ref{eq:linvz},\ref{eq:linvt}) into (\ref{eq:eleccurr}) and using $u=h=r^2$, $\chi=0$, the only non vanishing conductivity is the anomalous Hall
\be
\label{eq:anohallcond}
\sigma_{AH} = 8\alpha A_3(0)=8\alpha b_{IR}\,.
\ee
It is important to emphasize that, as it generally  happens with anomaly induced transport coefficients, the form of (\ref{eq:anohallcond}) is not modified by considering general  $\mathcal V,Z_1,Z_2,W$, showing the universality of the anomalous Hall conductivity. It is only necessary a non vanishing value of the axial field in the IR to have a non vanishing anomalous Hall conductivity, as expected. The other conductivities vanish, nonetheless we are interested in estimating their small $\omega$ dependence. In order to do so, it is necessary to change strategy and use the matching asymptotic technique.

\subsection{IR scaling of longitudinal and transverse conductivities}
If we plug the ansatz (\ref{eq:ansatzIR}) into the Eq. (\ref{eq:eqf3}), change coordinates and redefine fields as follow
\be
r= \frac{\omega}{u_0 x}  \qquad  , \qquad v_z = x^{\frac{3-\beta}{2}}p
\ee
the e.o.m. acquires the form of a Bessel equation
\be
\label{eq:longIR}
x^2p''(x)+x p'(x) + \left(x^2- \nu_1^2\right)p(x) = 0 \,,
\ee
with  $\nu_1 = \frac{3-\beta}{2}$. The solution satisfying the infalling condition is the Hankel function
\be
p(x) = H_{\nu_1}(x)\,.
\ee
 With this solution and following \cite{Charmousis:2010zz,Horowitz:2009ev,Gubser:2009ic} the small frequency behavior can be obtained
\be
\label{eq:condzz}
\re\, \sigma_{zz} \propto \omega^{2|\nu_1|-1} = \omega^{2-\beta}\,.
\ee

To solve Eq. (\ref{eq:dec1}) we notice that for the irrelevant solutions the Chern-Simons contribution in the equation is subleading whereas in the critical case it is not. Again changing variables
\be
 v_\pm = x^{\frac{1+\beta}{2}}q_\pm \,,
\ee
the equations read
\bea
\label{eq:transIR1}x^2q_\pm''(x)+x q_\pm'(x) + \left(x^2- 1\right)q_\pm(x) = 0 \quad&,& \beta=1\\
\label{eq:transIR2}x^2q_\pm''(x)+x q_\pm'(x) + \left(x^2\pm\frac{8 \alpha  \beta  b_{IR} }{Z_1^{IR}}x- \nu_2^2\right)q_\pm(x) = 0 \quad &,& \beta\neq 1
\eea
with $\nu_2=\frac{1+\beta}{2}$. The first equation is again the Bessel equation, and the associated conductivity  will be of the form (\ref{eq:condzz}). The second equation can be also solved analytically, and its infalling solution is
\be
q_\pm = e^{i x}x^{\nu_2}U\left(\nu_2+1/2 \mp i \frac{4 \alpha    b_{IR} }{Z_1^{IR}} \beta ,2\nu_2+1,-2 i x\right)\,,
\ee
where $U$ is the hypergeometric confluent function. Using again the method of matching asymptotes we find the scaling for the conductivity for the critical case, which turns out to be

\be
\re\, \sigma_{xx}= \re\, \sigma_{yy} \propto \omega^\beta\,.
\ee

Even when considering a general action of the type (\ref{eq:action2}), the general picture does not change much with respect to the observations of \cite{Landsteiner:2015pdh}. The anomalous Hall conductivity preserves its form, and it is not vanishing if and only if the time reversal breaking parameter $b$ flows in the IR to a non zero value. In the aforementioned paper the system shows a quantum phase transition when the parameter $M/b$ is suitably 
tuned, and the anomalous Hall conductivity plays the role of an order parameter.

\subsection{Full frequency dependence of conductivities}
\label{sec:FreqConduc}
Now we will turn to the computation of the full frequency dependence of the conductivities. To do so it is necessary to obtain the full $r-$dependence of the background fields, forcing us to select specific functions for $Z_1,Z_2,W$. We fix them to be $Z_1=Z_2=1$ and $W=q^2\chi^2$, considering that the qualitative behavior of the system is expected to be independent of their form.\footnote{Defining the functions in this way makes the action (\ref{eq:action2}) equivalent to the one used in \cite{Landsteiner:2015pdh}} We will also choose the same scalar potential as in \cite{Landsteiner:2015pdh}
\be\label{eq:scalarpot}
V(\chi) = -12- 3 \chi^2 + \frac{\lambda}{2}\chi^4\,,
\ee
with $q=\sqrt{3}$, $\lambda=15/8$.\footnote{We take those values for $q$ and $\lambda$ to simplify the numerical problem. For such $q$ and $\lambda$ the IR scaling dimensions for the scalar and axial operators $\Delta_{IR}$'s are integers, simplifying the near IR expansions.} We also fix the flavor number $N=4$
(Eq. \ref{eq:wardj5}).\footnote{ We choose this $N$ because TaAs \cite{Xu:2015vt} has precisely eight Weyl points close to the Fermi level, therefore the number of Dirac fermions needed is four. } 

Once all the free functions and the parameters in the action are fixed, the near IR fields reduce to:
\begin{figure}[t!]
	\begin{center}
		\includegraphics[width=0.8\textwidth]{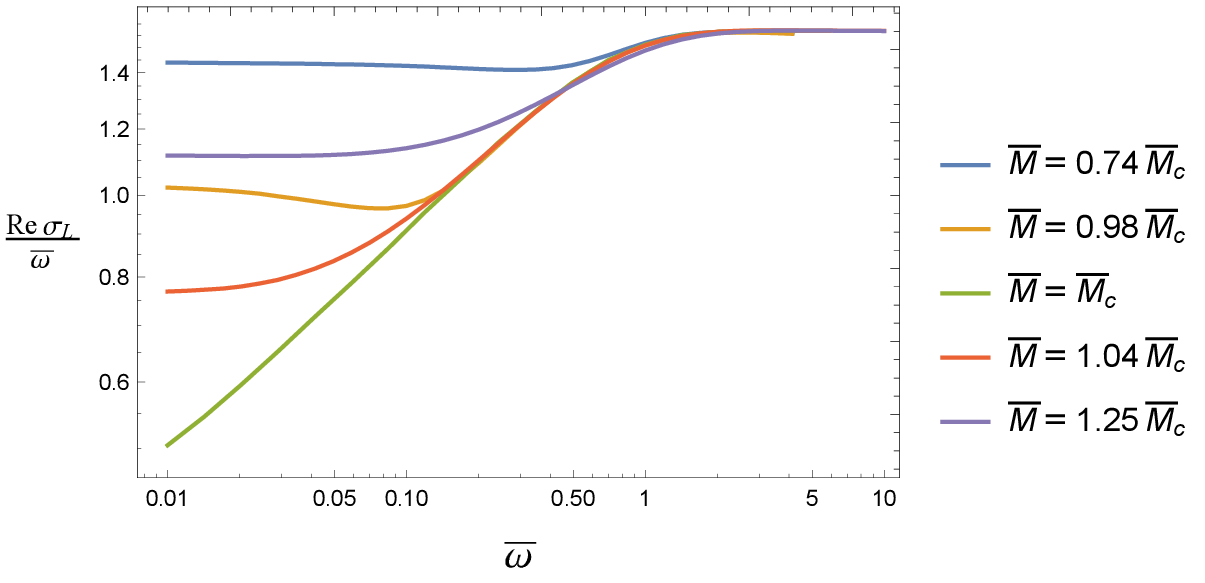}
		\caption{\small Real part of the longitudinal conductivity as a function of frequency for different values of $\bar M$ in the topological and trivial  semimetal phases, also the conductivity for the critical point is shown.}
		\label{fig_longCond}
	\end{center}
\end{figure}
\begin{itemize}
		\item Critical point
	\bea
	u(r) &=& u_0r^2 \left(1 + u_1 r^\alpha +\ldots\right)  \quad ,\quad h(r) = h_0r^{2\beta} \left(1+ h_1 r^\alpha +\ldots\right) \quad , \\
	A_z(r) &=& r^\beta \left(1 + a_1 r^\alpha +\ldots\right) \quad, \quad	\chi(r) = \chi_{IR} \left(1+ \chi_1 r^\alpha +\ldots\right)
	\eea
	where 
	\bea
	u_0 &\sim& 1.193 \quad ,\quad h_0 \sim 1.376 \quad , \quad \chi_{IR}\sim 0.661 \quad, \quad \beta \sim 0.733\\
	\nn	u_1 &\sim& 0.177\chi_1\quad ,\quad h_1 \sim -1.310\chi_1 \quad , \quad a_1 \sim 0.546\chi_1 \quad,\quad \alpha = 1.174\,.
	\eea
	After integrating out to boundary, the solution has the following ratio for the couplings
	\be
\bar M_c=	\frac{M_c}{b_c}  \sim 0.868\,.
	\ee

	From now on we will use the notation $\bar f$ to denote the quantity $f$ in units of $b$.
	\item Trivial phase
	\bea
	&&u(r) = u_0r^2+\ldots \quad ,\quad h(r) = h_0r^2 +\ldots\, ,\\
	&& A_z(r) = r^{\Delta_{b_{IR}}} +\ldots \quad, \quad	\chi(r) = \chi_{IR}(1 + \chi_1 r^{-4+\Delta_{\chi_{IR}}} +\ldots)
	\eea
	where 
	\be
	u_0 = \frac{6}{5}   \quad , \quad \chi_{IR}= 2  \sqrt{\frac{2}{5}} \quad,\quad	 \Delta_{b_{IR}}   = 2 \quad , \quad \Delta_{\chi_{IR}} =  5\,.
	\ee
	The two shooting parameters to build the full solution are ($h_0,\chi_1$), however the underlying conformal invariance of the system implies that the background will be a mono-parametric solution depending on $\bar M >\bar M_c$.
	\item Topological phase
	\bea
	&& u(r) = u_0r^2 +\ldots \quad ,\quad h(r) = h_0r^2 +\ldots \, , \\
	&& A_z(r) = 1 +\ldots \quad, \quad	\chi(r) =  \chi_1 r^{-3/2}e^{-s/r} +\ldots
	\eea
		where 
	\be
	u_0 = 1 \quad , \quad  s =  \sqrt{\frac{3}{h_0}}\,.
	\ee
This phase is characterized by having $\bar M<\bar M_c$. The shooting parameters are $h_0,\chi_1$.
	
\begin{figure}[t!]
	\begin{center}
		\includegraphics[width=0.43\textwidth]{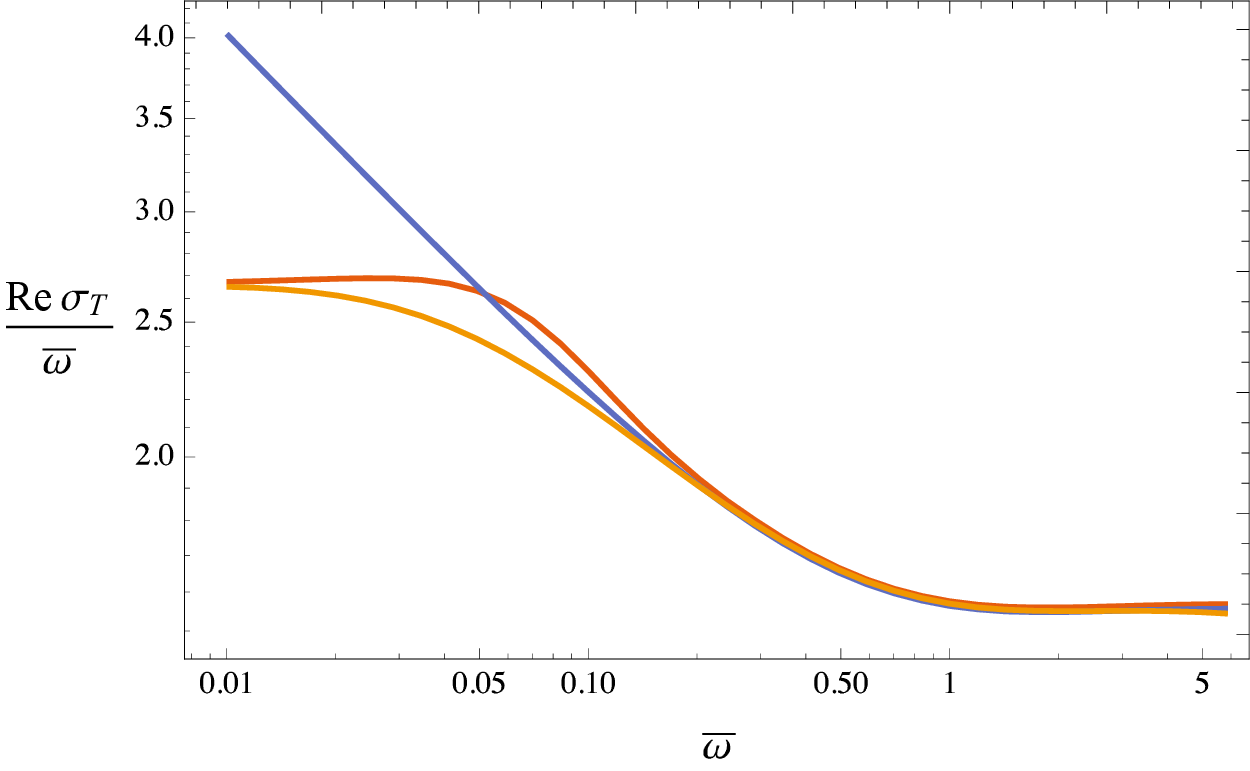}
		\includegraphics[width=0.56\textwidth]{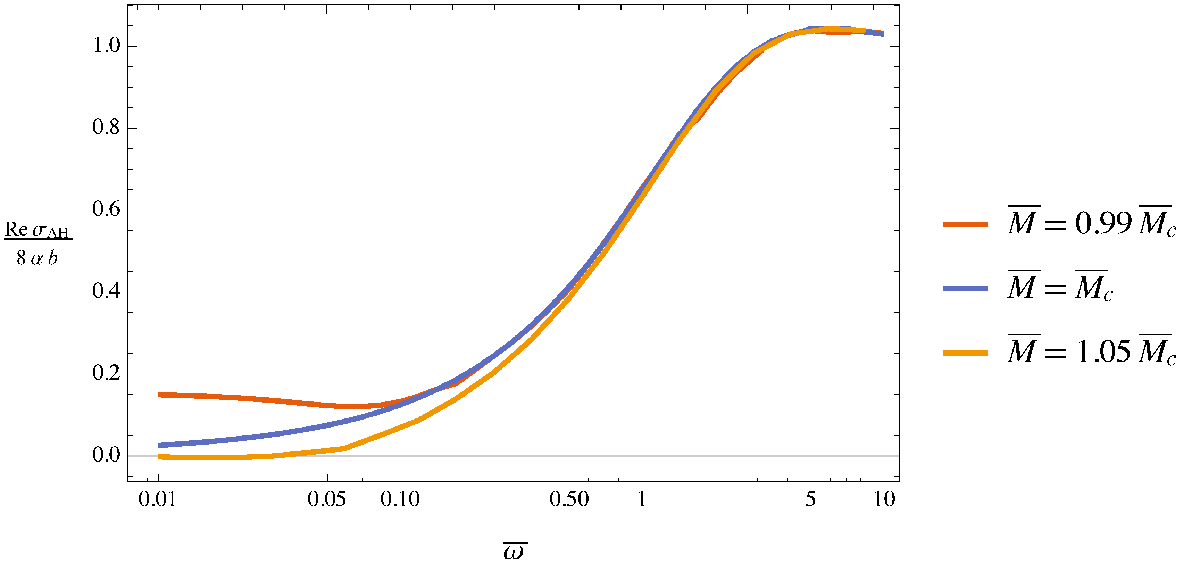}
		\caption{\small Left: Transverse conductivity as a function of frequency for the two different phases of the model and the critical point. Right: Anomalous Hall conductivity as a function of the frequency in the topological phase, the quantum critical point and in the trivial phase}
		\label{fig_transversCond}
	\end{center}
\end{figure}
\end{itemize}

After having constructed the background geometries we can proceed to solve the equations for fluctuations (\ref{eq:eomansatz1}-\ref{eq:eomansatz3}). To do so, we find a near IR expansion for the gauge field fluctuations and use them to integrate from the IR to the UV. Then, we plug them into (\ref{eq:eleccurr}) and extract the conductivities.

The first case to consider is the conductivity in the longitudinal sector, which is shown in Fig.~\ref{fig_longCond}. We compute the conductivity for critical solution (green) and we observe the expected behavior $\omega^{2-\beta}$. In the topological and trivial semimetal phases we observe, as expected, a linear frequency dependence at small frequencies.\footnote{Notice that the conductivity is divided by $\bar{\omega}$ to make clear the two different scalings, the linear and the critical ones.} However when $\bar M$ is close to the critical value in both phases we observe the emergence of an intermediate scaling given by the critical exponent $\beta$.

In Fig.~\ref{fig_transversCond} we show the frequency dependence of the transverse electrical  conductivity (left) and the anomalous Hall conductivity (right). We have computed the conductivities for $\bar M\sim\bar M_c$ in the topological and trivial phases besides the critical $\bar M_c$ case. As also observed in Fig.~\ref{fig_longCond}, both phases show a linear conductivity in the IR and UV, and they agree for some intermediate regime with the critical conductivity. The conductivity in the critical phase shows a scaling exponent that agrees very well with the predicted value $\omega^\beta$. In the right plot,  we observe how the anomalous Hall conductivity approaches zero in the critical and trivial phases, unlike the topological phase in which the DC conductivity is nonzero. At high frequencies the anomalous Hall conductivity always takes the value $\sigma_{AH}^{(UV)}=8\alpha b$. This result is not unexpected considering that when the energies are high enough the system should behave as massless, and axial symmetry must be restored (modulo the anomaly breaking term).

\begin{figure}[t!]
	\begin{center}
		\includegraphics[width=0.8\textwidth]{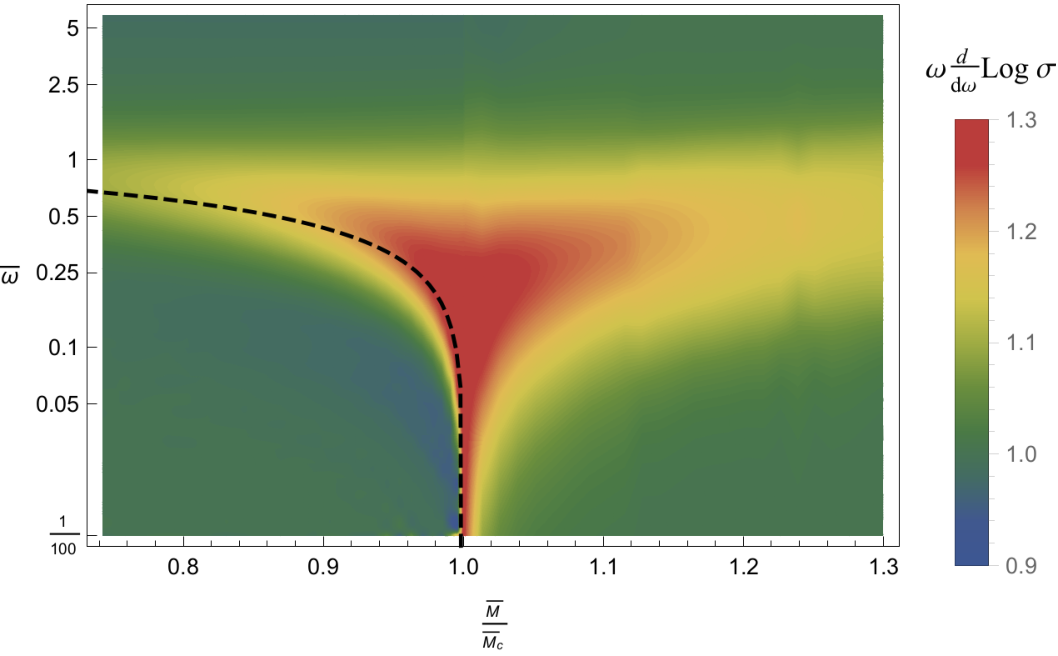}
		\caption{\small Reconstruction of the quantum phase transition by computing the regions in which the conductivity shows a powerlaw. In the vertical axes we plot the energy scale, given by the frequency, and in the horizontal axes the mass parameter $\bar M$. The green regions correspond with a linear conductivity and the
			red zone to the quantum critical region where the power law is determined by exponent of the quantum critical point $\beta$.}
		\label{fig_phaseDiagram}
	\end{center}
\end{figure}

Considering the observation of emergence of the critical scaling in the conductivity for $\bar M\sim\bar M_c$, we computed the conductivity as a function of ($\bar{\omega},\bar{M}$), in order to sketch the quantum phase diagram of the system using the AC conductivity of the material\footnote{In \cite{Roy:jy} the optical conductivity was also used to reconstruct the quantum phase diagram on a disordered Weyl Semimetal.}. In Fig.~\ref{fig_phaseDiagram} we show in a contour plot the quantity
\be
m = \omega\frac{d}{d\omega}\log\sigma_L\,,
\ee
which gives the exponent of the conductivity  within the regions where it shows a power-law, otherwise it is a meaningless $\omega$-dependent function.\footnote{We only analyse the longitudinal conductivity, the transverse conductivity should reproduce qualitatively  similar results.} Notice that in the regions where $m$ is not a constant we observe in the figure a gradient in the colors. In the plot we observe the presence of the quantum critical point and a well defined  quantum critical region (red area), extending up to some orders of magnitude above $\omega=0$. We could expect to reproduce the phase diagram of the system with this computation, due to the underlying scale invariance of the system. At zero temperature the conductivity has to be a function of the form $\sigma(\omega,b,M)=f(\bar{\omega},\bar{M})$. On the other hand, at zero frequency but finite temperature, the energy scale is given by $\bar T$. Therefore for the finite $T$ and $\omega=0$ case, the power law in frequency observed at $T=0$ has to be preserved interchanging $\omega\leftrightarrow T$.

The phase diagram shows four well defined regions that can be understood in terms of the physical scales of the problem.

First we have the UV region (upper  green) when the energies are large, \mbox{$\omega \gg b$}.
The second region corresponding to the Weyl semimetal phase (left green) is manifest at energies smaller than the physical separation of the Weyl cones $\omega \ll  b_{IR}$. In the figure we also included a dashed black line representing the separation of the cones as a function of the mass parameter, and we observe how it determines the transition region between the Weyl semimetal and the quantum critical phases. The trivial semimetal phase (right green region) is characterized by energies much smaller than the UV cut-off. 
In summary, in order for the quantum critical region to be manifest in the conductivity, the separation of the Weyl nodes has to be much smaller than the UV cut-off scale.

\section{Phenomenological implications}
\label{sec:expe}
\begin{figure}[t!]
	\begin{center}
		\includegraphics[width=0.48\textwidth]{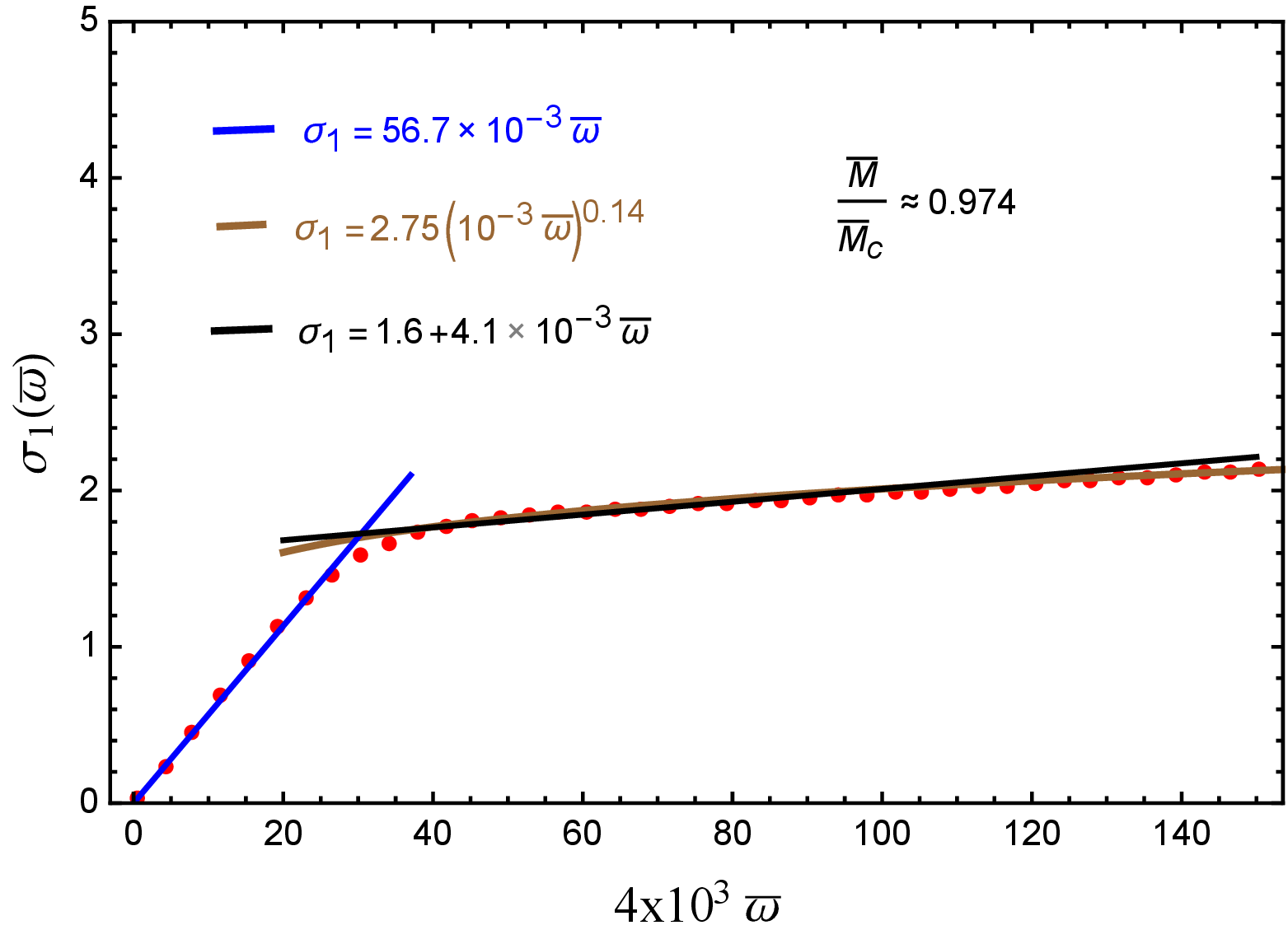}
	\includegraphics[width=0.49\textwidth]{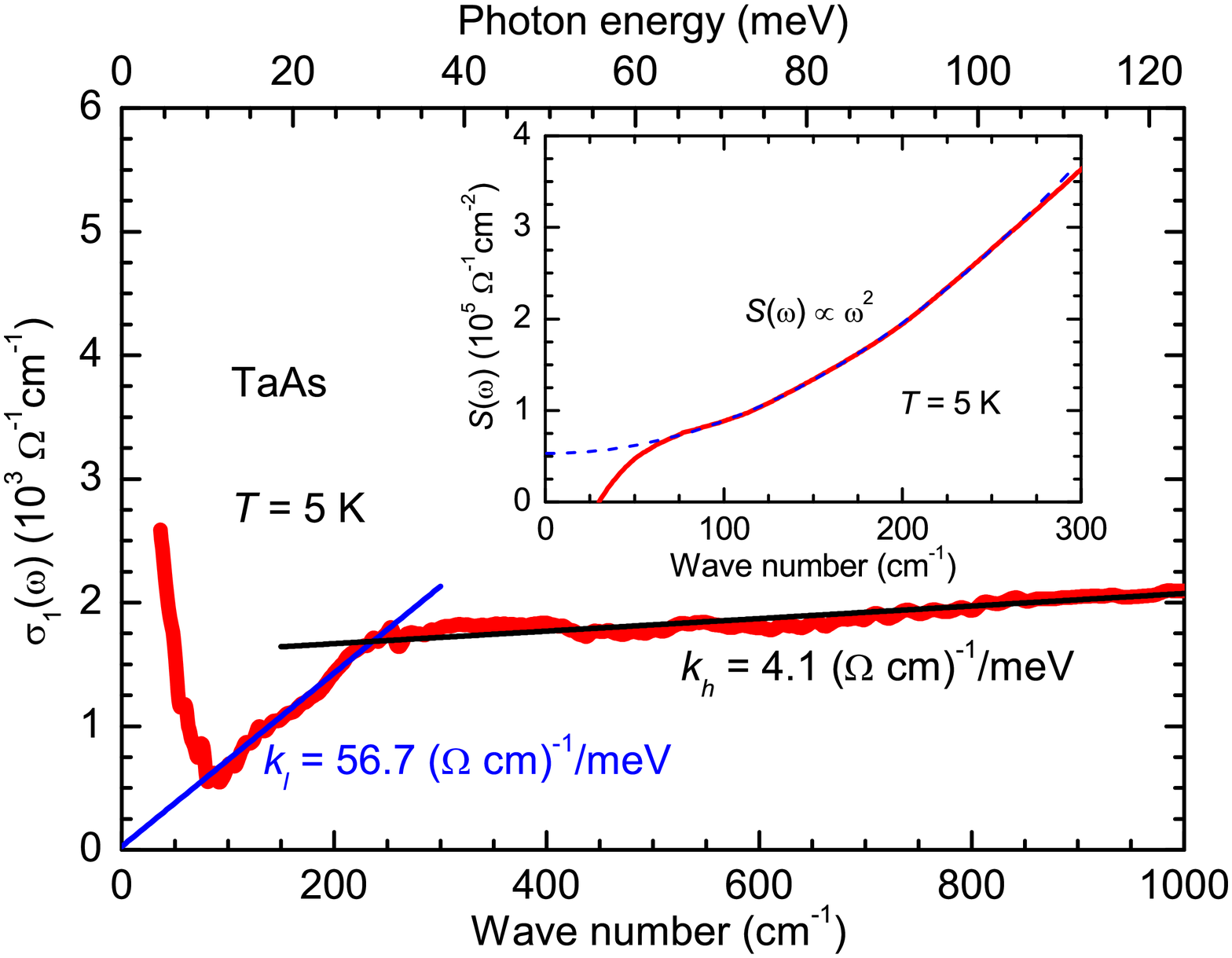}
			\caption{\small Left plot: Transverse conductivity as a function of the conductivity ($\sigma_1=10.7\re\,\sigma_T$). Right Plot: Experimental data (image taken from \protect\cite{Xu:2015vt}), $\sigma_1$ is the real part of the conductivity along the (107) surface of the BZ. In the left plot blue and black lines correspond with the fitting used in the right plot. Brown line is the fitting using the dynamical exponent $\beta=0.14$.}
		\label{fig_exp}
	\end{center}
\end{figure}

 A recent experiment \cite{Xu:2015vt} measured the optical conductivity of a recently discovered Weyl semimetal  (TaAs), in particular when 
 the temperature is $T=5K$ three differentiated regimes were observed (see right plot of Fig.~\ref{fig_exp}):
\begin{itemize}
	\item A Drude peak up to frequencies $\omega\sim 10\,$meV.
	\item Between $\omega\sim 10-30\,$meV a linear dependence with a slope of $56.7$($\Omega\,$ cm)$^{-1}$/meV.
	\item Between $\omega\sim 30-120\,$meV a linear dependence with a slope of $4.1$($\Omega\,$ cm)$^{-1}$/meV.
\end{itemize}

Actually, the aspects that concern to us are the two last items. A Drude peak would have been appeared in the holographic model after switching on temperature, introducing chemical potential and breaking translational invariance.  
Nonetheless our goal is to understand the power-laws of the conductivity and 
the mechanism that leads to the change in the slope at $\omega \simeq 30$~meV. This last aspect is possibly the most interesting one,
since the small frequency behavior ($\omega \lesssim 30$~meV) is already well described by the predictions of the weak coupling computations \cite{Hosur:2011fl,Burkov:2011de}, while a satisfactory description of the high energy regime is still missing.
Considering that $\omega\gg T$ is the region of interest, we can expect to describe the physics with a zero temperature computation. 
Moreover, the persistence of a linear behavior in the conductivity up to 120~meV suggests that in this regime
the physics of the system is dominated by the low energy linear dispersion relation around the Weyl points. 
Actually this  condition is  necessary in order for the IR physics of our holographic description to be applicable to the measurement; we shall
further comment on this point below.

Theoretical models in agreement with experiments showed the presence of twelve pairs of Weyl points close to the Fermi energy in the band structure of TaAs \cite{Weng:2015aa,huang_weyl_2015}. Four pairs, denoted as W1, lie 2~meV above the Fermi energy. The remaining eight pairs (W2) are 21~meV
below the Fermi energy. For small enough frequencies only the physics close to the W1 points is relevant since the interband transitions near the 
W2 points require energies of at least 42~meV. A possible explanation of the change in the slope in the linear behavior of the conductivity for 
$\omega \gtrsim 30$~meV could then be related to the fact that at such energies also the physics near the W2 points starts to contribute.
However this interpretation leads to a tension between the fitting of the Fermi velocity from the data using Eq.~\eqref{eq:condweak}
and the reasonable range of values expected for the Fermi velocity derived both theoretically and experimentally. 

\begin{figure}[t!]
	\begin{center}
		\includegraphics[width=0.7\textwidth]{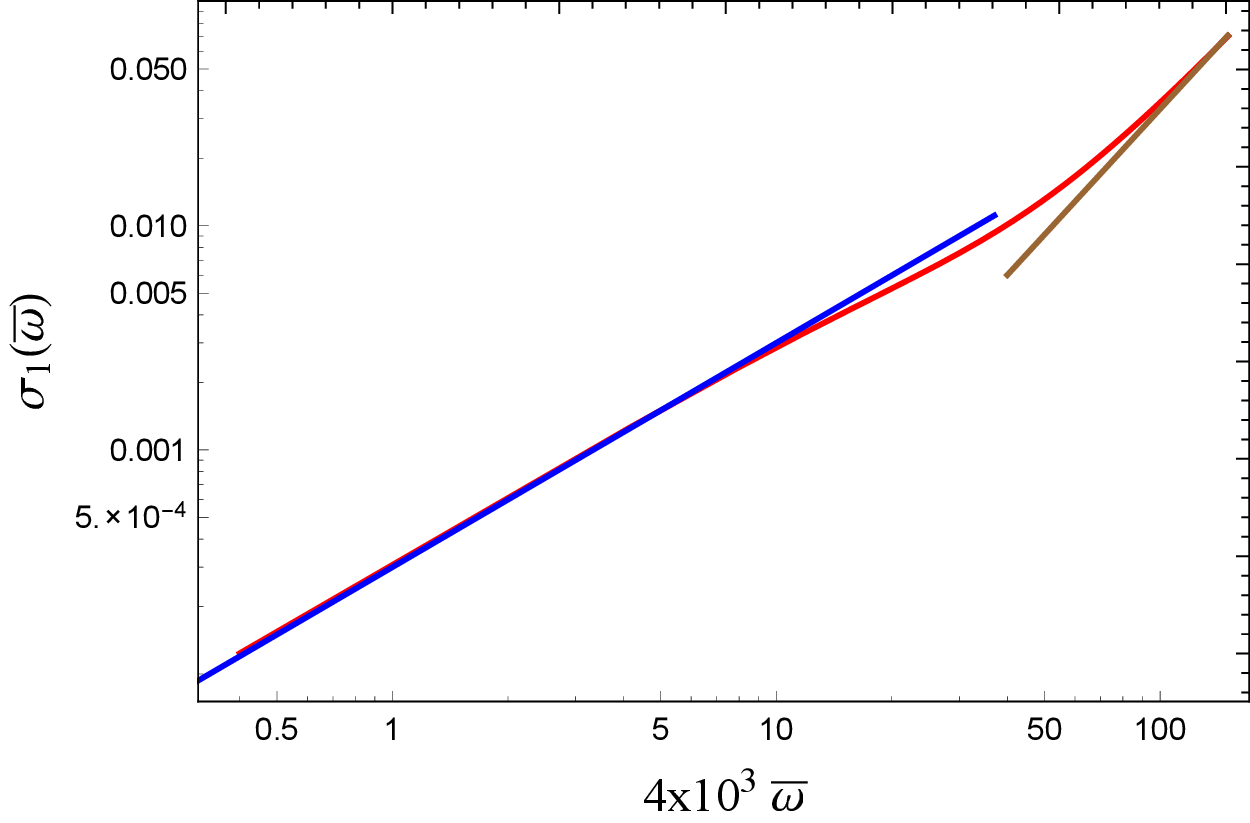}
		\caption{\small Real part of the longitudinal conductivity (red line)  as a function of frequency. Blue line corresponds with a linear fitting and brown line with the critical exponent $\omega^{1.86}$. ($\sigma_1=10.7\re\,\sigma_L$)}
		\label{fig_expL}
	\end{center}
\end{figure}

Motivated by the results of our model, we were tempted to propose a different explanation for the higher frequencies regime 
of the optical conductivity. We will consider the possibility that it is determined by the quantum critical region of the phase diagram of the material. 
As observed in Fig.~\ref{fig_phaseDiagram} the scaling in conductivity is different from the linear behavior for energies sitting inside the quantum critical region, i.e. energies higher than the cones separation, but smaller than the UV cut-off. In particular, within our framework the conductivity scaling in the quantum critical region is given by
\be
\sigma_L \sim \omega^{2-\beta}\qquad,\qquad\sigma_T \sim \omega^\beta\,,
\ee
and  $\beta$ will take always values between $(0,1)$. If we choose $q=49/100$ and $\lambda=1/30$, the critical exponent and mass are
\bea
\beta &\approx& 0.14 \quad\implies \sigma_T\sim\omega^{0.14}\,,\quad \sigma_L\sim\omega^{1.86}\,,\\
 \bar M_c &\approx& 0.664 .
\eea
With this critical exponent it is difficult to distinguish, in the transverse conductivity, between an (almost horizontal) straight line and $\omega^{0.14}$ as we show in Fig.~\ref{fig_exp}. In the left plot of the aforementioned figure we observe the transverse conductivity for $\bar M= 0.647$, noticing a great similarity with the experimental conductivity along the (107) surface of the Brillouin zone (BZ) of the material  (right plot). On the other side, the longitudinal conductivity has a complete different behavior as can be seen in Fig.~\ref{fig_expL}, where the conductivity is almost parabolic. Another important feature is that the longitudinal conductivity is two orders of magnitude smaller than the transverse one, because $\omega^{1.86}$ at small frequencies decays much faster than $\omega^{0.14}$.  However, due to the higher amount of Weyl points in the real material
and their distribution in momentum space, the same notion of transverse and longitudinal conductivity is absent. In a more realistic holographic setup (more than two unaligned Weyl nodes) the conductivity will not split into transverse and longitudinal, but the effects of the quantum critical point will remain. A more realistic model needs to be studied in detail, in order to check whether the new critical exponent would be compatible with the experimental data.

Although our model has to be simply understood as a toy model for real Weyl semimetals, the results it provides suggest 
a novel explanation for the high-frequency behavior of the optical conductivity.
According to the latter the change in the slope for TaAs 
at energies $\sim 30$~meV would not be determined by the physics near the W2 points, 
but rather by the occurrence of the transition from the Weyl semimetal phase to the critical region. 
Nevertheless it is worth noticing that in order to trust this interpretation we need to understand  the hierarchy of scales in the system. 
As seen in Fig. \ref{fig_phaseDiagram} the scale at which the system enters in the critical region is set by the separation of the Weyl points in the momentum space.  
If this scale happens to be much larger than the energy at which the interband transitions near the W2 nodes turn on (42~meV)
we obviously cannot rely on this explanation.
However, the theoretically predicted values for the separation of Weyl nodes W1 and W2 respectively are \cite{huang_weyl_2015}
\be
b_1\sim 0.03\;\mathring{\mathrm{A}}^{-1} \qquad ,\qquad b_{2} \sim  0.07\;\mathring{\mathrm{A}}^{-1},
\ee
which via the dispersion relation $\omega = v_f k$, and after using $v^{(1)}_f\sim 0.3-1.7$~eV$\mathring{\mathrm{A}}$, $v^{(2)}_f\sim 0.2-2.4$~eV$\mathring{\mathrm{A}}$ \cite{Huang:2015gy,Lv:kp} give the energy scales
\be
E_{b_1}\sim 8-40\; \mathrm{meV}\qquad , \qquad E_{b_{2}}\sim 14-160\; \mathrm{meV}\,,
\ee
energies which roughly sit within the frequencies studied in the experiment, and comparable also to the energy scale of
the interband transitions near the W2. This fact does not rule out the possibility of having a contribution from the quantum critical region of the system on the second slope. Of course this conclusion has to be taken with great care. The holographic model we have used takes only into account the presence of two Weyl cones, whereas the physical system has the set of W1 and W2 nodes with different separations. This fact will imply that conductivities will not be decomposed in term of longitudinal and transverse respect to the cone's separation, and the critical exponents may be modified in a more realistic model. However, the presence of a quantum phase transition will remain, and the effects of the quantum critical region on the optical conductivity shall be present.

Besides the separation of the Weyl points with opposite chirality within  W1 or W2, which, as we argued, defines the energy scale associated to the 
transition to quantum critical region, another relevant scale in the system is given by the separation between pairs of W1 with the ones of W2  (see Figure~3
in \cite{huang_weyl_2015}). 
This distance roughly determines the UV cut-off of the system where
the low energy description (relativistic Weyl fermions) ceases to be valid, since it sets the scale where the deviations from the linearity in the dispersion relation appear.\footnote{The reader may wonder why we use this separation and not $b_1$ and $b_2$ as an estimate for the cut-off. The reason is that within the model the separation of points with opposite chirality sets the transition in the power-law, even though at this scale there is  already a deviation from linearity.}
From \cite{Huang:2015gy} it turns out that this separation is one order of magnitude bigger than $b_2$, supporting the explanation for the second slope suggested by the model. 

A clear way of testing our proposal would be to tune the analog of $\bar{M}$ in the material and to measure the conductivity. In such an experiment, if the second slope is associated to the quantum critical region, the farther the system is from the quantum critical point the higher the transition scale will be.  Unfortunately in TaAs it is
not clear how to control experimentally the phase transition \cite{Chang:2016hp}. Recently,  another material 
showing a tunable type II Weyl semimetal phase ($\chem{Mo}_x \chem{W}_{1-x}\chem{Te}_2$) has been predicted \cite{Chang:2016hp} and experimentally discovered \cite{Belopolski:2016es}. It would be highly interesting to see what is the behavior of the conductivity in such a type of material.


\section{Conclusions}
\label{sec:concl}
We have generalized the study of \cite{Landsteiner:2015pdh} analysing  the possibility of having Weyl semimetal phases in holography by using the general action 
\bea\label{eq:actionc}
\nn S_0 &=& \int \dd^5x \sqrt{-g} \left[ \, R- \mathcal V(\chi)  - \frac{1}{4} Z_1(\chi)H^2 - \frac{1}{4}Z_2(\chi)F^2 - (\partial \chi)^2 + W(\chi)A^2\right.\\
\label{eq:actionconc}&& \left.+\frac{\alpha}{3}\epsilon^{M N R P Q} A_{M} \big( F_{NR} F_{PQ} + 3 H_{NR}H_{PQ}\big)\right] \, ,
\eea
within this framework we have proved that Weyl semimetal phases are allowed if the gauge field is massless in the IR, the scalar field runs to a constant value and simultaneously extremizes the scalar potential and gauge field mass function. Another general result is the existence of a quantum critical point with a Lifshitz-like scaling symmery and dynamical exponent $\beta$, which always takes values between zero and one. We also observed that if the scalar runs to a constant value, the quantum phase transition will be always between the Weyl semimetal phase and a topologically trivial semimetal.
After the classification of the IR fixed points we computed the optical conductivity of the system in all the phases obtaining a linear conductivity in the IR as expected by dimensional analysis, however the presence of the quantum critical point in the phase diagram introduces a quantum critical region in which the conductivity has a scaling given by the dynamical exponent $\beta$. The time-reversal breaking parameter breaks the isotropy of the space-time leading in the IR  to an anisotropic conductivity given by 
\bea
\sigma_L &\sim& \omega^{2-\beta}\,,\\
\sigma_T &\sim& \omega^{\beta}\,.
\eea
 We also verified that the anomalous Hall conductivity obeys an universal form
 \be
 \sigma_{AH} = 8\alpha b_{IR}\,,
 \ee
 with $b_{IR}$ being the renormalized time-reversal parameter, which we interpreted as the effective separation of Weyl cones. The universality of the result does not come as a surprise considering that it is intimately related with the anomaly, and it is well known that this type of transport coefficients are protected.

We reconstructed the phase diagram of the theory by computing the power-laws exponents as a function of $\bar{M}$. Remarkably enough the  phase diagram  showed the standard features of a quantum phase diagram, with a quantum critical region that extends several orders of magnitude above the zero energy case.

Finally we compared our results with the experiment \cite{Xu:2015vt}. The experimental data has been qualitatively reproduced within our model after setting the parameters in such a way of having a quantum critical region, characterized by a small critical exponent $\beta=0.14$. Whether the physics determining the second slope in the right plot of Fig.~\ref{fig_exp} is given by the quantum critical region or not can be verified by properly measuring the matrix of conductivities  and  varying the coupling $\bar M$, this tuning would shift the transition scale depending on how far from the critical point the system is.

In the future it would be  worth studying a more realistic phase diagram. To do so it would be necessary to consider IR geometries with a logarithmic running for the scalar field, allowing the system to have different IR fixed points, like Lifshitz and hyperscaling violating Lifshitz geometries, which may be insulating. It would be also interesting the inclusion of extra gauge fields in order to model the presence of more than two Weyl nodes.

\begin{acknowledgments}
We would like to thank Elias Kiritsis, Karl Landsteiner and Yan Liu, for enlightening discussions.  F. PB. would like also to acknowledge the IFT-UAM/CSIC for the warm hospitality during his visit while the development of this paper. The authors are grateful to the Galileo Galilei Institute for theoretical physics, for the hospitality when finishing the present work.
\end{acknowledgments}

\appendix

\section{Gibbons-Hawking action and Counterterm}
\label{app:CT}
In this appendix we show the explicit form of the Gibbons-Hawking boundary action,
\bea
S_{GH} &=& \int_{r=\Lambda} \dd^4x\,\sqrt{-\gamma}\,2 K\,,
\eea
and the counterterm needed to renormalize the theory in the case of having $Z_1=Z_2=Z_3=1$ and the scalar potential Eq. (\ref{eq:scalarpot})
\bea
\nn S_{CT} &=& \int_{r=\Lambda} \dd^4x\,\sqrt{-\gamma}\left( \frac{1}{2}\mathcal V_{UV} - |\phi|^2 + \log r \left[ \frac{1}{4}F^2+\frac{1}{4}H^2 + |D_\mu\phi|^2 + \left(\frac{1}{3} + \frac{\lambda}{2} \right)|\phi|^4 \right] \right)\,,\\
\eea
where $\gamma_{\mu\nu}$ is the  induced boudary metric and $K$  the trace of the extrisic curvature.

\section{Equations of motion}
\label{app:EOM}
The equations of motion for the action (\ref{eq:action2}) can be written as follow 

\bea
\label{eq:eom1}
&&R_{ M N} + \frac{Z_2}{2} F_{ M R} {F^{ R}}_{ N} + \frac{Z_1}{2} H_{ M R} {H^{ R}}_{ N} -  \partial_{M} \chi \partial_{N} \chi + W A_M A_N +\nn \\
&& \frac{g_{ M N}}{2} \Big[(\partial_{M} \chi )^{2} + \mathcal V - R + \frac{Z_2}{4}F^2 + \frac{Z_1}{4}H^2 - W A^2 \Big] = 0 \, ,
\eea

\begin{equation}
\label{eq:eom2}
\frac{1}{\sqrt{-g}} \, \partial _{ M} \big( \sqrt{-g} \,Z_1 H^{ N M} \big) - 2 \alpha \, \epsilon^{ N M R P Q} F_{ M R} H_{ P Q} =0 \, ,
\end{equation}

\begin{equation}\label{eq:eom3}
\frac{1}{\sqrt{-g}}\partial _{ M} \big( \sqrt{-g} Z_2F^{ N M} \big) -  \alpha \epsilon^{ N M R P Q}  \Big[ F_{ M R} F_{ P Q} + H_{ M R} H_{ P Q} \Big] - 2WA^N =0 \, ,
\end{equation}

\begin{equation}
\label{eq:eom4}
\frac{1}{\sqrt{-g}} \, \partial_{ M} \big( \sqrt{-g} \, \partial^{ M} \chi \big) = \frac{1}{8}\partial_\chi Z_1 H^2 +\frac{1}{8}\partial_\chi Z_2 F^2 + \frac{1}{2}\partial_\chi W A^2+ \frac{1}{2} \partial_\chi \mathcal V\,.
\end{equation}

After plugging the ansatz (\ref{eq:ansatz}) into the previous equations we obtain for the gravity sector
\bea
\label{eq:eomansatz1}
\frac{ Z_2A_3^{'2}}{2 h(r)}+\frac{h''}{2 h}-\frac{h^{'2}}{4 h^2}+\frac{u''}{u}-\frac{u^{'2}}{2 u^2}+\chi ^{'2} &=& 0  \,,\\
-\frac{Z_2A_3^{'2} }{4 h}+\frac{A_3^2 W}{2 h u}+\frac{3u'}{4u}\left(\frac{ h' }{ h }+\frac{u'}{ u}\right)+\frac{\mathcal V}{2 u}-\frac{1}{2} \chi ^{'2} &=& 0\,,\\
\frac{Z_2A_3^{'2} }{2 h}+\frac{WA_3^2 }{ h u}+\frac{u'}{2u}\left(\frac{3 h' }{2 h }-\frac{u'}{ u}\right) -\frac{h^{'2}}{4 h^2}+\frac{h''}{2 h}-\frac{u''}{2 u}&=& 0 \,,
\eea
only two of them are linearly independent. The gauge sector reads
\be
\label{eq:eomansatz2}
\sqrt{h}\left(\frac{u^2  Z_2A_3'}{\sqrt{h}} \right)' - 2u WA_z = 0\,,
\ee
and the scalar
\be
\label{eq:eomansatz3}
\frac{1}{\sqrt{h} u}\left(\sqrt{h} u^2 \chi '\right)' -\partial_\chi Z_2\frac{u   A_3^{'2}}{4 h}-\partial_\chi W\frac{  A_3^2}{2 h}-\frac{1}{2}   \partial_\chi \mathcal V = 0\, .
\ee

\section{IR Perturbations}
\label{app:IRPert}

Assuming the following ansatz for the fields
\bea
\dd s^2 &=& u(r)(-\dd t^2 + \dd x_1^2+\dd x_2^2)  + h(r)\dd x_3^2 +\frac{\dd r^2}{u(r)}\,,\\
A &=& A_3(r)\dd x_3 \,,\\
\chi &=& \chi(r) \,,
\eea
the solutions can be split into a leading and subleading contribution in the IR
\bea
&&u(r) = u_0 r^{2}\left(1+\delta u(r)+\ldots\right) \qquad,\qquad h(r) = h_0 r^{2\beta}\left(1+\delta h(r)+\ldots\right) \,, \\
&& A_3(r) = r^c( b_{IR}+\delta A_r(r)+\ldots) \qquad,\qquad \chi(r) = \chi_{IR}+\delta \chi(r) +\ldots.
\eea
The solutions for the subleading IR corrections can be obtained for each phase of the model.
\begin{itemize}
	\item {\bf Weyl Semimetal Phase ($b_{IR}=1,\, c=0$) :} Corrections to the Weyl semimetal phase are of the following type
	
	\be
	\delta u = \frac{c_1}{r}+\frac{c_2}{r^4} \, ,\qquad	\delta h = \frac{c_1}{r}-\frac{2c_2}{r^4}+c_3 \, ,\qquad	\delta A_3 = c_4+ \frac{c_5}{ r^2} \, ,\qquad
	\ee
	$c_3,c_4$ are marginal deformations that just redefine the value of $h_0$ and $b_{IR}$ respectively and can be scaled out. The rest of the modes are relevant and will destroy the AdS IR geometry.
	
	The only perturbation with an irrelevant deformation is the scalar field
	\be
		\delta \chi = c_6 r^{-3/2} e^{-\frac{s}{r}} +c_7 r^{-3/2} e^{\frac{s}{r}}\,,
		\ee
	where $s = \sqrt{1/2W_{IR}''h_0^{-1}}L_{IR}$. If $W_{IR}''<0$ any perturbation would destroy the IR. On the other hand if $W_{IR}''>0$ the perturbation associated to $c_6$ would be irrelevant in the IR.
	
	\item {\bf Trivial Semimetal Phase  ($b_{IR}=0,\, c=0$):} Corrections to the trivial semimetal phase are of the following type
	\be
	\delta u = \frac{c_1}{r}+\frac{c_2}{r^4} \, ,\qquad	\delta h = \frac{c_1}{r}-\frac{2c_2}{r^4}+c_3 \,.
	\ee
	These perturbations have the same form as in the previous case, and have to be all set to zero. However the scalar and gauge field have irrelevant modes 
	\be
		\delta A_3 = c_4 r^{-2 - \Delta_{b_{IR}} }+ c_5 r^{\Delta_{b_{IR}}} \,,
	\ee
	where $\Delta_{b_{IR}} = -1+\sqrt{1 + \frac{2W_{IR}L_{IR}^2}{ Z_2^{IR}}}$. $c_4$ corresponds to a relevant deformation, but $c_5$ is irrelevant. With the scalar field something similar happens; one mode is relevant and the other one irrelevant
	\be
	\delta \chi = c_6 r^{ - \Delta_{\chi_{IR}} }+ c_7 r^{-4+\Delta_{\chi_{IR}}}\, ,
	\ee
	where $\Delta_{\chi_{IR}} = 2+\sqrt{4 + m_{IR}^2L_{IR}^2}$. Stability implies $m_{IR}^2L_{IR}^2>0$. We are using here the definition $\mathcal V_{IR}''=2m_{IR}^2$.
	
		\item {\bf Critical Point ($b_{IR}=1,\, c=\beta$):} In the critical case, the zero temperature deformation to the Lifshitz IR takes the following form
		
		\be
		\delta u =u_1 r^\alpha \,,\qquad	\delta h = h_1 r^\alpha \,,\qquad	\delta A_3 = a_1 r^\alpha \,,\qquad \delta \chi = \chi_{IR}\chi_1 r^\alpha \,,
		\ee
		where $\chi_1$ is the only free constant. The rest is
		\be
		u_1 = \frac{2 (\beta -1)  \chi_{IR} \left(3 \beta ^2 u_0 Z_{IR}'-2(\alpha +2 \beta -1) W_{IR}'\right)}{3 (\alpha +1) \beta  u_0 \left(\alpha ^2+\alpha  (\beta +3)-2 (\beta -3) (\beta -1)\right) Z_{IR}}\chi_1
		\ee
			\bea
	\nn	h_1 &=& \frac{8   \chi_{IR} (\beta -1) \left(\alpha ^2+3 \alpha  \beta +\alpha +\beta  (\beta +2)\right) W_{IR}'}{3 \alpha  (\alpha +1) \beta  u_0 \left(\alpha ^2+\alpha  (\beta +3)-2 (\beta -3) (\beta -1)\right) Z_{IR}}\chi_1 +\\
		&&-\frac{6  \chi_{IR}  \beta ^2 u_0 (\beta -1) (\alpha  (\alpha -\beta +7)+6) Z_{IR}'}{3 \alpha  (\alpha +1) \beta  u_0 \left(\alpha ^2+\alpha  (\beta +3)-2 (\beta -3) (\beta -1)\right) Z_{IR}}\chi_1
		\eea
			\bea
	\nn	a_1 &=&\frac{2 \chi_{IR} \left(3 \alpha ^2+\alpha  (6 \beta -3)+2 \left(\beta ^2+\beta -2\right)\right) W_{IR}'}{3 \alpha  (\alpha +1) u_0 \left(\alpha ^2+\alpha  (\beta +3)-2 (\beta -3) (\beta -1)\right) Z(\chi_{IR})} \chi_1+\\
		&& -\frac{\beta  \chi_{IR} (\alpha  (\alpha  (\alpha +4)-(\beta -7) \beta -3)+6 (\beta -1)) Z_{IR}'}{\alpha  (\alpha +1) \left(\alpha ^2+\alpha  (\beta +3)-2 (\beta -3) (\beta -1)\right) Z_{IR}} \chi_1\,.
		\eea
		The exponent $\alpha$ can be obtained inverting the following equation
		\bea
\nn	\mathcal	V_{IR}'' &=& 2 \alpha  u_0 (\alpha +\beta +3) + \frac{(\beta -1) \beta  u_0 Z_{IR}''}{Z_{IR}}+\frac{2(\beta -1) W_{IR}''}{\beta  Z(\chi_{IR})}+\\
\nn	&& \frac{4 (\beta -1) (\beta +2) W_{IR}^{'2}}{3 \beta ^2 u_0 \left(\alpha ^2+\alpha  (\beta +3)-2 (\beta -3) (\beta -1)\right) Z_{IR}^2} +\\
\nn	&& \frac{4 (\beta -1) (3 \beta -5) W_{IR}' Z_{IR}'}{\left(\alpha ^2+\alpha  (\beta +3)-2 (\beta -3) (\beta -1)\right) Z_{IR}^2} +\\
&& -\frac{2 (\beta -1) \beta  u_0 \left(\alpha ^2+\alpha  (\beta +3)+\beta  (11-2 \beta )-6\right) Z_{IR}^{'2}}{\left(\alpha ^2+\alpha  (\beta +3)-2 (\beta -3) (\beta -1)\right) Z_{IR}^2}\,.
		\eea
		The previous polynomial has four possible solutions for $\alpha$ and only two of them may be real numbers, depending on the values of the other parameters.
\end{itemize}

\bibliographystyle{JHEP} 
\bibliography{WeylConduc}

\end{document}